\documentclass[article]{jss}\usepackage[]{graphicx}\usepackage[]{xcolor}
\makeatletter
\def\maxwidth{ %
  \ifdim\Gin@nat@width>\linewidth
    \linewidth
  \else
    \Gin@nat@width
  \fi
}
\makeatother

\definecolor{fgcolor}{rgb}{0.345, 0.345, 0.345}
\newcommand{\hlnum}[1]{\textcolor[rgb]{0.686,0.059,0.569}{#1}}%
\newcommand{\hlstr}[1]{\textcolor[rgb]{0.192,0.494,0.8}{#1}}%
\newcommand{\hlopt}[1]{\textcolor[rgb]{0,0,0}{#1}}%
\newcommand{\hlstd}[1]{\textcolor[rgb]{0.345,0.345,0.345}{#1}}%
\newcommand{\hlkwa}[1]{\textcolor[rgb]{0.161,0.373,0.58}{\textbf{#1}}}%
\newcommand{\hlkwb}[1]{\textcolor[rgb]{0.69,0.353,0.396}{#1}}%
\newcommand{\hlkwc}[1]{\textcolor[rgb]{0.333,0.667,0.333}{#1}}%
\newcommand{\hlkwd}[1]{\textcolor[rgb]{0.737,0.353,0.396}{\textbf{#1}}}%

\usepackage{framed}
\makeatletter
\newenvironment{kframe}{%
 \def\at@end@of@kframe{}%
 \ifinner\ifhmode%
  \def\at@end@of@kframe{\end{minipage}}%
  \begin{minipage}{\columnwidth}%
 \fi\fi%
 \def\FrameCommand##1{\hskip\@totalleftmargin \hskip-\fboxsep
 \colorbox{shadecolor}{##1}\hskip-\fboxsep
     \hskip-\linewidth \hskip-\@totalleftmargin \hskip\columnwidth}%
 \MakeFramed {\advance\hsize-\width
   \@totalleftmargin\z@ \linewidth\hsize
   \@setminipage}}%
 {\par\unskip\endMakeFramed%
 \at@end@of@kframe}
\makeatother

\definecolor{shadecolor}{rgb}{.97, .97, .97}
\definecolor{messagecolor}{rgb}{0, 0, 0}
\definecolor{warningcolor}{rgb}{1, 0, 1}
\definecolor{errorcolor}{rgb}{1, 0, 0}
\newenvironment{knitrout}{}{} 

\usepackage{alltt}

\usepackage{orcidlink,thumbpdf,lmodern}

\usepackage{framed}

\usepackage{amsmath, amssymb}

\author{Lineu Alberto Cavazani de Freitas~\orcidlink{0000-0002-0076-6642}\\Paraná Federal University
   \And Wagner Hugo Bonat~\orcidlink{0000-0002-0349-7054}\\Paraná Federal University}
\Plainauthor{Lineu Alberto Cavazani de Freitas, Wagner Hugo Bonat}

\title{Hypothesis tests for multiple responses regression models in \proglang{R}: The \pkg{htmcglm} Package}

\Plaintitle{Hypothesis tests for multiple responses regression models in R: The htmcglm Package}

\Shorttitle{htmcglm: hypothesis tests for multiple responses regression models}

\Abstract{

This article describes the \proglang{R} package \pkg{htmcglm} implemented for performing hypothesis tests on regression and dispersion parameters of multivariate covariance generalized linear models (McGLMs). McGLMs provide a general statistical modeling framework for normal and non-normal multivariate data analysis along with a wide range of correlation structures. The proposed package considers the Wald statistics to perform general hypothesis tests and build tailored ANOVAs, MANOVAs and multiple comparison tests. The goal of the package is to provide tools to improve the interpretation of regression and dispersion parameters. We assess the effects of the covariates on the response variables by testing the regression coefficients. Similarly, we perform tests on the dispersion coefficients in order to assess the correlation between study units. It could be of interest in situations where the data observations are correlated with each other, such as in longitudinal, times series, spatial and repeated measures studies. The \pkg{htmcglm} package provides a user friendly interface to perform MANOVA like tests as well as multivariate hypothesis tests for models of the mcglm class. We describe the package implementation and illustrate it through the analysis of two data sets. The first deals with an experiment on soybean yield; the problem has three response variables of different types (continuous, counting and binomial) and three explanatory variables (amount of water, fertilization and block). The second dataset addresses a problem where responses are longitudinal bivariate counts of hunting animals; the explanatory variables used are the hunting method and sex of the animal. With these examples we were able to illustrate several tests in which the proposal proves to be useful for the evaluation of regression and dispersion parameters both in problems with dependent or independent observations.

}


\Plainkeywords{multivariate regression models, McGLM, hypothesis tests, Wald test, ANOVA, MANOVA, Multiple comparisons, R}

\Address{
  Lineu Alberto Cavazani de Freitas\\
  Department of Informatics\\
  Paraná Federal University\\
  Centro Politécnico\\
  Curitiba 81531980, CP 19081, Paraná, Brazil.\\
  E-mail: \email{lineuacf@gmail.com}\\
}
\IfFileExists{upquote.sty}{\usepackage{upquote}}{}
\begin{document}


\section{Introduction} \label{sec:intro}

The \pkg{htmcglm} package for \proglang{R} \citep{R2022} provides functions for performing hypothesis testing on parameters of multivariate covariance generalized linear models (McGLMs; \citet{Bonat16}) as fitted by the \pkg{mcglm} package \citep{mcglm}. 

McGLMs provide a general statistical modeling framework for normal and non-normal multivariate data analysis along with a wide range of correlation structures. 
McGLMS are specified by a set of regression, dispersion, power and correlation parameters. Each set of parameters has a very useful practical interpretation.

By analysing the regression parameters, it is possible to assess the effect of the explanatory variables on the response variables. 
Simililarly, by analysing the dispersion parameters, we can assess the correlation structure between study units.
It is useful in situations where the observations of the data set are correlated with each other, such as in longitudinal, times series and repeated measures studies. 
The power parameters provide us an indication of which probability distribution could fit well to the response variable. Finally, the correlation parameters measure 
the strength of the association between response variables in a multivariate context.

The development of hypothesis tests for the purpose of evaluating these quantities is of great interest in practical problems and leads to procedural forms for evaluating the resulting quantities of the model. 
The \pkg{htmcglm} package is a full \proglang{R} implementation with functions based on Wald statistics to evaluate regression and dispersion parameters. 
The features include functions for general linear hypothesis testing, univariate and multivariate analysis of variance tables, as well as multivariate multiple comparison tests.

The \pkg{htmcglm} package is available from the Comprehensive R Archive Network (CRAN) at \href{https://cran.r-project.org/package=htmcglm}{https://cran.r-project.org/package=htmcglm} and complement the functions available in the \pkg{mcglm} package \citep{mcglm}.

There are several implementations of the Wald test in different contexts in \proglang{R}. The package \pkg{lmtest} \citep{lmtest} has a generic function to perform Wald tests to compare nested linear and generalized linear models. The package \pkg{survey} \citep{survey1,survey2,survey3} has a function that performs Wald tests that, by default, tests whether all coefficients associated with a given regression term are zero, but it is possible to specify hypotheses with other values.

The package \pkg{car} \citep{car} has an implementation to test linear hypotheses about parameters of linear models, generalized linear models, multivariate linear models, mixed effects models, among others; in this implementation, the user has full control of which parameters to test and with which values to compare in the null hypothesis.

For analysis of variance tables, \proglang{R} has the function \code{anova()} in the standard package \pkg{stats} \citep{R2022} applicable to linear and generalized linear models. The package \pkg{car} \citep{car} has a function that returns analysis of variance tables of types II and III for different models. For multiple comparisons, one of the main packages available is \pkg{multcomp} \citep{multcomp} which provides an interface for testing multiple comparisons for parametric models.

However, when dealing with multivariate covariance generalized linear models fitted in the \pkg{mcglm} package, there is only one type of analysis of variance implemented in the library and there are no options for performing general linear hypothesis tests, nor multiple comparison tests. Therefore, as it is a flexible class of models with high application potential, our goal is to provide computational implementation of hypothesis tests for McGLMs in such a way that it is possible to test general linear hypotheses, generate analysis of variance tables, multivariate analysis of variance tables and multiple comparisons tests.

The article is organized as follows. In \autoref{sec:mcglm} we present a review of the general structure and estimation of the parameters of a McGLM, based on the ideas of \citet{Bonat16}. In \autoref{sec:wald} the details of the Wald test to evaluate assumptions about parameters of a McGLM are presented. In \autoref{sec:implementacao} we introduce the \proglang{R} implementation discussing the main functions available in the \pkg{htmcglm} package. We illustrate the package use through some examples in \autoref{sec:exemplos}. Finally, \autoref{sec:conclusao} presents a discussion and directions for future work on the improvement of the \pkg{htmcglm} package.


\section{Multivariate covariance generalized linear models}\label{sec:mcglm}

In this section, we present a review of the McGLMs specification. Consider $\boldsymbol{Y}_{N \times R} = \left \{ \boldsymbol{Y}_1, \dots, \boldsymbol{Y}_R \right \}$ is a matrix of response variables and $\boldsymbol{M}_{N \times R} = \left \{ \boldsymbol{\mu}_1, \dots, \boldsymbol{\mu}_R \right \}$ is a matrix of expected values. The variance and covariance matrix for each response $r$, $r = 1,..., R$, is denoted by $\Sigma_r$ and has dimension $N \times N$. In addition, consider an $R \times R$ correlation matrix, denoted by $\Sigma_b$, to describe the correlation between the response variables. The McGLMs \citep{Bonat16} are specified by:

$$
      \begin{aligned}
        \mathrm{E}(\boldsymbol{Y}) &=
          \boldsymbol{M} =
            \{g_1^{-1}(\boldsymbol{X}_1 \boldsymbol{\beta}_1),
            \ldots,
            g_R^{-1}(\boldsymbol{X}_R \boldsymbol{\beta}_R)\}
          \\
        \mathrm{Var}(\boldsymbol{Y}) &=
          \boldsymbol{C} =
            \boldsymbol{\Sigma}_R \overset{G} \otimes
            \boldsymbol{\Sigma}_b,
      \end{aligned}
$$

\noindent where the functions $g_r()$ are standard link functions; $\boldsymbol{X}_r$ denotes a $N \times k_r$ design matrix; $\boldsymbol{\beta}_r$ denotes a $k_r \times 1$ vector of regression parameters. $\boldsymbol{\Sigma}_R \overset{G} \otimes \boldsymbol{\Sigma}_b = \mathrm{Bdiag}(\tilde{\boldsymbol{\Sigma}}_1, \ldots, \tilde{\boldsymbol{\Sigma}}_R) (\boldsymbol{\Sigma}_b \otimes \boldsymbol{I}) \mathrm{Bdiag}(\tilde{\boldsymbol{\Sigma}}_1^\top, \ldots, \tilde{\boldsymbol{\Sigma}}_R^\top)$ is the Generalized Kronecker product \citep{martinez13}. The matrix $\tilde{\boldsymbol{\Sigma}}_r$ denotes the lower triangular matrix of the Cholesky decomposition of the matrix ${\boldsymbol{\Sigma}}_r$. The operator $\mathrm{Bdiag()}$ denotes the block-diagonal matrix and $\boldsymbol{I}$ is a $N \times N$ identity matrix. 

For continuous, binary, binomial and bounded data, the variance and covariance matrix $\boldsymbol{\Sigma}_r$ is given by:

$$
\Sigma_r =
\mathrm{V}\left(\boldsymbol{\mu}_r; p_r\right)^{1/2}(\boldsymbol{\Omega}\left(\boldsymbol{\tau}_r\right))\mathrm{V}\left(\boldsymbol{\mu}_r; p_r\right)^{1/2}.
$$

In the case of count response variables, the variance and covariance matrix for each response variable is given by:

$$
\Sigma_r = diag(\boldsymbol{\mu}_r)+ \mathrm{V}\left(\boldsymbol{\mu}_r; p_r\right)^{1/2}(\boldsymbol{\Omega}\left(\boldsymbol{\tau}_r\right))\mathrm{V}\left(\boldsymbol{\mu}_r; p_r\right)^{1/2},
$$

\noindent where $\mathrm{V}\left(\boldsymbol{\mu}_r; p_r\right) = diag(\vartheta(\boldsymbol{\mu}_r; p_r))$ denotes a diagonal matrix in which the entries are given by the variance function $\vartheta(\cdot; p_r)$ applied elementwise to the vector $\boldsymbol{\mu}_r$. Different choices of variance functions $\vartheta(\cdot; p_r)$ imply different assumptions about the distribution of the response variable. We mention three options of variance functions: the Tweedie variance function, the Poisson-Tweedie dispersion function and the binomial variance function.

The Tweedie variance function characterizes the Tweedie family of distributions, is given by $\vartheta\left(\cdot; p_r\right) = \mu^{p_r}_r$, in which some distributions stand out: Normal ($p$ = 0), Poisson ($p$ = 1), gama ($p$ = 2) and inverse Gaussian ($p$ = 3) \citep{Jorgensen87, Jorgensen97}. 

The Poisson-Tweedie dispersion function \citep{Jorgensen15} is indicated for events defined by counts. It is given by $\vartheta\left(\cdot; p\right) = \mu + \tau\mu^p$ where $\tau$ is the dispersion parameter. Thus, we have a rich class of models for dealing with responses that characterize counts, since many important distributions appear as special cases, such as: Hermite ($p$ = 0), Neyman type A ($p$ = 1), negative binomial ($p$ = 2) and Poisson–inverse Gaussian (p = $3$).

Finally, the binomial variance function, given by $\vartheta\left(\cdot; p_r\right) = \mu^{p_{r1}}_r(1 - \mu_r)^{p_{r2}}$ can deal with binary, binomial and continuous bounded respose variables.
It is possible to notice that the power parameter $p$ appears in all the variance functions discussed. This parameter is important because it is an index that distinguishes between different 
probability distributions. Thus, it can be regarded as a tool for probability distribution automatic selection.

The dispersion matrix $\boldsymbol{\Omega({\tau})}$ describes the part of the covariance within each response variable that does not depend on the mean structure, that is, the correlation structure between the observations in the sample. Based on the ideas of \citet{Anderson73} and \citet{Pourahmadi00}, \citet{Bonat16} proposed to model the dispersion matrix through a matrix linear predictor combined with a covariance link function given by:

$$
h\left \{ \boldsymbol{\Omega}(\boldsymbol{\tau}_r) \right \} = \tau_{r0}Z_0 + \ldots + \tau_{rD}Z_D,
$$

\noindent where $h()$ is the covariance link function, $Z_{rd}$ with $d$ = 0,$\ldots$, $D$ are matrices that represent the covariance structure for each response variable $r$ and $\boldsymbol{\tau_r}$ = $(\tau_{r0}, \ldots, \tau_{rD})$ is a $(D + 1) \times 1$ vector of dispersion parameters. 

Some possible covariance link functions are identity, inverse and exponential-matrix. The specification of the covariance link function is discussed by \citet{Pinheiro96} and it is possible to select combinations of matrices to obtain the some well-known models in the literature for longitudinal data, time series, spatial and spatio-temporal data. Further details are discussed by \citet{Demidenko13}.

Thus, the McGLMs configure a general framework for analysis via regression models for Gaussian and non-Gaussian data with multiple responses, in which no assumptions are made regarding the independence of the observations. The class is defined by three functions (link, variance and covariance), in addition to a linear predictor and a matrix linear predictor for each response under analysis. 

\subsection{Estimation and inference}\label{sec:inf}

McGLMs are fitted based on the estimating function approach described in detail by \citet{Bonat16} and \citet{jorg04}. This subsection presents an overview of the algorithm and the asymptotic distribution of the estimators based on estimating functions.

McGLM's second-moment assumptions allow us to split the parameter vector into two subsets: $\boldsymbol{\theta} = (\boldsymbol{\beta}^{\top}, \boldsymbol{\lambda}^{\top})^{\top}$. Thus, $\boldsymbol{\beta} = (\boldsymbol{\beta}_1^\top, \ldots, \boldsymbol{\beta}_R^\top)^\top$ is a  $K \times 1$ vector of regression parameters and $\boldsymbol{\lambda} = (\rho_1, \ldots, \rho_{R(R-1)/2}, p_1, \ldots, p_R, \boldsymbol{\tau}_1^\top, \ldots, \boldsymbol{\tau}_R^\top)^\top$ is a $Q \times 1$ vector of dispersion parameters. Furthermore, $\mathcal{Y} = (\boldsymbol{Y}_1^\top, \ldots, \boldsymbol{Y}_R^\top)^\top$ denotes the $NR \times 1$ stacked response variables vector. Similarly, $\mathcal{M} = (\boldsymbol{\mu}_1^\top, \ldots, \boldsymbol{\mu}_R^\top)^\top$ denotes the $NR \times 1$ stacked expected values vector. 

In order to estimate the regression parameters, the quasi-score function \citep{Liang86} is given by 

$$
\begin{aligned}
  \psi_{\boldsymbol{\beta}}(\boldsymbol{\beta},
  \boldsymbol{\lambda}) = \boldsymbol{D}^\top
  \boldsymbol{C}^{-1}(\mathcal{Y} - \mathcal{M}),
\end{aligned}
$$

\noindent where $\boldsymbol{D} = \nabla_{\boldsymbol{\beta}} \mathcal{M}$ is a $NR \times K$ matrix and $\nabla_{\boldsymbol{\beta}}$ denotes the gradient operator. The $K \times K$ sensitivity matrix of $\psi_{\boldsymbol{\beta}}$ is given by

$$
\begin{aligned}
S_{\boldsymbol{\beta}} = E(\nabla_{\boldsymbol{\beta} \psi \boldsymbol{\beta}}) = -\boldsymbol{D}^{\top} \boldsymbol{C}^{-1} \boldsymbol{D},
\end{aligned}
$$

\noindent whereas the $K \times K$ variability matrix of $\psi_{\boldsymbol{\beta}}$ is written as

$$
\begin{aligned}
V_{\boldsymbol{\beta}} = VAR(\psi \boldsymbol{\beta}) = \boldsymbol{D}^{\top} \boldsymbol{C}^{-1} \boldsymbol{D}.
\end{aligned}
$$

For the dispersion parameters, the following Pearson estimating function is adopted,

$$
  \begin{aligned}
    \psi_{\boldsymbol{\lambda}_i}(\boldsymbol{\beta},
    \boldsymbol{\lambda}) =
    \mathrm{tr}(W_{\boldsymbol{\lambda}i}
    (\boldsymbol{r}^\top\boldsymbol{r} -
    \boldsymbol{C})),  i = 1,.., Q, 
  \end{aligned}
$$

\noindent where $W_{\boldsymbol{\lambda}i} = -\frac{\partial \boldsymbol{C}^{-1}}{\partial \boldsymbol{\lambda}_i}$ and $\boldsymbol{r} = (\mathcal{Y} - \mathcal{M})$. The entry $(i,j)$ of the $Q \times Q$ sensitivity matrix of $\psi_{\boldsymbol{\lambda}}$ is given by

$$
  \begin{aligned}
    S_{\boldsymbol{\lambda_{ij}}} = E \left (\frac{\partial }{\partial \boldsymbol{\lambda_{i}}} \psi \boldsymbol{\lambda_{j}}\right) = -tr(W_{\boldsymbol{\lambda_{i}}} CW_{\boldsymbol{\lambda_{J}}} C).
  \end{aligned}
$$

\noindent The entry $(i,j)$ of the $Q \times Q$ variability matrix of $\psi_{\boldsymbol{\lambda}}$ is defined by

$$
  \begin{aligned}
V_{\boldsymbol{\lambda_{ij}}} = Cov\left ( \psi_{\boldsymbol{\lambda_{i}}}, \psi_{\boldsymbol{\lambda_{j}}} \right) = 2tr(W_{\boldsymbol{\lambda_{i}}} CW_{\boldsymbol{\lambda_{J}}} C) + \sum_{l=1}^{NR} k_{l}^{(4)} (W_{\boldsymbol{\lambda_{i}}})_{ll} (W_{\boldsymbol{\lambda_{j}}})_{ll},
  \end{aligned}
$$

\noindent where $k_{l}^{(4)}$ denotes the fourth cumulant of $\mathcal{Y}_{l}$. In the McGLM estimation process its empirical version is used.

To take into account the covariance between the vectors $\boldsymbol{\beta}$ and $\boldsymbol{\lambda}$, \citet{Bonat16} obtained the cross-sensitivity and variability matrices, denoted by $S_{\boldsymbol{\lambda \beta}}$, $S_{\boldsymbol{\beta \lambda}}$ and $V_{\boldsymbol{\lambda \beta}}$, more details, see \citet{Bonat16}. The joint sensitivity and variability matrices of $\psi_{\boldsymbol{\beta}}$ and $\psi_{\boldsymbol{\lambda}}$ are denoted by

$$
  \begin{aligned}
    S_{\boldsymbol{\theta}} = \begin{bmatrix}
      S_{\boldsymbol{\beta}} & S_{\boldsymbol{\beta\lambda}} \\ 
      S_{\boldsymbol{\lambda\beta}} & S_{\boldsymbol{\lambda}} 
      \end{bmatrix} \text{e } V_{\boldsymbol{\theta}} = \begin{bmatrix}
      V_{\boldsymbol{\beta}} & V^{\top}_{\boldsymbol{\lambda\beta}} \\ 
      V_{\boldsymbol{\lambda\beta}} & V_{\boldsymbol{\lambda}} 
    \end{bmatrix}.
  \end{aligned}
$$

Let $\boldsymbol{\hat{\theta}} = (\boldsymbol{\hat{\beta}^{\top}}, \boldsymbol{\hat{\lambda}^{\top}})^{\top}$ be the estimating functions estimators of $\boldsymbol{\theta}$. Then, the asymptotic distribution of $\boldsymbol{\hat{\theta}}$ is

$$
  \begin{aligned}
    \boldsymbol{\hat{\theta}} \sim N(\boldsymbol{\theta}, J_{\boldsymbol{\theta}}^{-1}),
  \end{aligned}
$$

\noindent where $J_{\boldsymbol{\theta}}^{-1}$ is the inverse of the Godambe information matrix, given by $J_{\boldsymbol{\theta}}^{-1} = S_{\boldsymbol{\theta}}^{-1} V_{\boldsymbol{\theta}} S_{\boldsymbol{\theta}}^{-\top}$, where $S_{\boldsymbol{\theta}}^{-\top} = (S_{\boldsymbol{\theta}}^{-1})^{\top}.$

To solve the system of equations $\psi_{\boldsymbol{\beta}} = 0$ and $\psi_{\boldsymbol{\lambda}} = 0$ the following modified Chaser algorithm is adopted 
$$
\begin{aligned}
\begin{matrix}
\boldsymbol{\beta}^{(i+1)} = \boldsymbol{\beta}^{(i)}- S_{\boldsymbol{\beta}}^{-1} \psi \boldsymbol{\beta} (\boldsymbol{\beta}^{(i)}, \boldsymbol{\lambda}^{(i)}), \\ 
\boldsymbol{\lambda}^{(i+1)} = \boldsymbol{\lambda}^{(i)}\alpha S_{\boldsymbol{\lambda}}^{-1} \psi \boldsymbol{\lambda} (\boldsymbol{\beta}^{(i+1)}, \boldsymbol{\lambda}^{(i)}).
\end{matrix}
\end{aligned}
$$

The procedure aforementioned is implements in the \pkg{mcglm} package \citep{mcglm}.


\section{Wald Test for McGLMs}\label{sec:wald}

Following the ideas of \citet{htmcglm_proposta}, let $\boldsymbol{\theta^{*}}$ be a $h \times 1$ vector of parameters disregarding the correlation parameters and $J^{\boldsymbol{*}-1}$ the corresponding inverse of the Godambe information matrix. Let $\boldsymbol{L}$ be a $s \times h$ hypotheses specification matrix and $\boldsymbol{c}$ a $s \times 1$ vector with the values under the null hypothesis. In this notation, $s$ represents the number of restrictions. The hypotheses to be tested can be written as:

\begin{equation}
\label{eq:hipoteses_wald}
H_0: \boldsymbol{L}\boldsymbol{\theta^{*}} = \boldsymbol{c} \ vs \ H_1: \boldsymbol{L}\boldsymbol{\theta^{*}} \neq \boldsymbol{c}. 
\end{equation}

\noindent Thus, the generalization of the Wald test statistic to verify the validity of a hypothesis about parameters of a McGLM is given by:

$$
W = (\boldsymbol{L\hat\theta^{*}} - \boldsymbol{c})^T \ (\boldsymbol{L \ J^{\boldsymbol{*}-1} \ L^T})^{-1} \ (\boldsymbol{L\hat\theta^{*}} - \boldsymbol{c}),
$$

\noindent where $W \sim \chi^2_s$, that is, regardless of the number of parameters in the hypotheses, the test statistic $W$ is a single value that asymptotically follows the $\chi^2$ distribution with degrees of freedom given by the number of constraints, that is, the number of rows in the matrix $\boldsymbol{L}$, denoted by $s$.

In general, each column of the matrix $\boldsymbol{L}$ corresponds to one of the $h$ parameters of $\boldsymbol{\theta^{*}}$ and each row to a constraint. Its specification basically consists of filling the matrix with $0$, $1$ and eventually $-1$ in such a way that the product $\boldsymbol{L}\boldsymbol{\theta^{*}}$ correctly represents the hypotheses of interest. The correct specification of $\boldsymbol{L}$ allows us testing any linear hipothesis for each parameter individually or even formulating hypotheses for several parameters jointly.

\citet{htmcglm_proposta} presents examples of how to test different types of hypotheses of interest that arise in practical contexts. In this article, we shall present two examples: hypotheses for multiple parameters and hypotheses considering regression or dispersion parameters for responses under the same linear predictor.

For purposes of illustration, consider the case in which one wants to investigate whether a numeric variable $X_1$ has an effect on two response variables. Let $Y_1$ and $Y_2$ denote the response variables. A bivariate McGLM for this problem may have a linear predictor given by:

\begin{equation}
\label{eq:pred_ex}
g_r(\mu_r) = \beta_{r0} + \beta_{r1} X_1, r=1,2,
\end{equation}

\noindent where the index $r$ denotes the response variable, $r = 1,2$; $\beta_{r0}$ represents the intercept; $\beta_{r1}$ the regression coefficient associated with the variable $X_1$. We assume that each response has only one dispersion parameter $\tau_{r0}$ and that the power parameters were fixed. Therefore, it is a problem in which there are two response variables and only one explanatory variable. Further, we assume that the observations are independent, so $Z_0 = I$. 

Suppose the interest is to assess whether there is sufficient evidence to state that there is an effect of the explanatory variable $X_1$ on both response variables simultaneously. 
In this cas, we  have to test two parameters: $\beta_{11}$, which associates $X_1$ with the first response variable; and $\beta_{21}$, which associates $X_1$ with the second response variable. 
We can write the hypothesis as follows:

\begin{equation}
\label{eq:ex2}
H_0: \beta_{r1} = 0 \ vs \ H_1: \beta_{r1} \neq 0,
\end{equation}

\noindent or, equivalently:

$$
H_0: 
\begin{pmatrix}
\beta_{11} \\ 
\beta_{21}
\end{pmatrix} 
= 
\begin{pmatrix}
0 \\ 
0
\end{pmatrix}
\ vs \ 
H_1: 
\begin{pmatrix}
\beta_{11} \\ 
\beta_{21}
\end{pmatrix} 
\neq
\begin{pmatrix}
0 \\ 
0 
\end{pmatrix}.
$$

The hypotheses in the form of \autoref{eq:hipoteses_wald} have the following elements:

\begin{itemize}
  
  \item $\boldsymbol{\theta^{*T}}$ = $\begin{bmatrix} \beta_{10} \  \beta_{11} \ \beta_{20} \ \beta_{21} \ \tau_{11} \ \tau_{21} \end{bmatrix}$.

\item $\boldsymbol{L} = \begin{bmatrix} 0 & 1 & 0 & 0 & 0 & 0 \\
0 & 0 & 0 & 1 & 0 & 0 \end{bmatrix}.$
 
\item $\boldsymbol{c} = \begin{bmatrix} 0 \\ 0 \end{bmatrix}.$ 

\end{itemize}

The vector $\boldsymbol{\theta^{*}}$ has six elements and the matrix $\boldsymbol{L}$ has six columns. In this case, we are testing two parameters, so the matrix $\boldsymbol{L}$ has two rows. These lines are composed of zeros, except in the columns referring to the parameter of interest. It is simple to verify that the product $\boldsymbol{L}\boldsymbol{\theta^{*}}$ represents the hypothesis of interest, see Eq. \autoref{eq:ex2}. 
Thus, the asymptotic distribution of the test is $\chi^2_2$.

The \autoref{eq:pred_ex} describes a generic bivariate model. It is important to note that in this example both responses are subject to the same predictor. In practice, when it comes to McGLMs, different predictors can be specified between response variables. However, in cases where the responses are subject to identical predictors and the hypothesis about the parameters do not change from response to response, an alternative specification of the procedure is to use the Kronecker product to test the same hypothesis on multiple responses as used in \citet{plastica}.

Suppose that, in this example, the hypotheses of interest are still written as in the form of \autoref{eq:ex2}. However, as this is a bivariate model with the same predictor for the two responses, the hypothesis of interest are the same between responses and involves only regression parameters. Consequently, it is convenient to write the matrix $\boldsymbol{L}$ as the Kronecker product of two matrices: a matrix $\boldsymbol{G}$ and a matrix $\boldsymbol{F}$, ie, $\boldsymbol{L}$ = $\boldsymbol{G} \otimes \boldsymbol{F}$. In this way, the matrix $\boldsymbol{G}$ has dimension $R \times R$ and specifies the hypotheses about the responses, whereas the matrix $\boldsymbol{F}$ specifies the hypotheses between variables and has dimension ${s}' \times {h}'$, where ${s}'$ is the number of linear constraints, that is, the number of parameters tested for a single response, and ${h}'$ is the total number of coefficients of regression or dispersion of the response. Therefore, the matrix $\boldsymbol{L}$ has dimension (${s}'R \times h$).

In general, the matrix $\boldsymbol{G}$ is an identity matrix with a dimension equal to the number of responses composing the model. Whereas the matrix $\boldsymbol{F}$ is equivalent to a matrix $\boldsymbol{L}$ if there was only a single response in the model and only regression or dispersion parameters. We use the Kronecker product of these two matrices to ensure that the hypothesis described in the $\boldsymbol{F}$ matrix will be tested on $R$ model responses.

Thus, considering that this is the case in which the hypotheses can be rewritten by decomposing the $\boldsymbol{L}$ matrix, the test elements are given by:

\begin{itemize}
  
  \item $\boldsymbol{\beta^{T}}$ = $\begin{bmatrix} \beta_{10} \  \beta_{11} \  \beta_{20} \  \beta_{21} \end{bmatrix}$: the model regression parameters.

\item $\boldsymbol{G} = \begin{bmatrix} 1 & 0 \\ 0 & 1  \end{bmatrix}$: identity matrix with dimension given by the number of responses.

\item $\boldsymbol{F} = \begin{bmatrix} 0 & 1 \end{bmatrix}$: equivalent to a $\boldsymbol{L}$ for a single response.

\item $\boldsymbol{L} = \boldsymbol{G} \otimes \boldsymbol{F} =  \begin{bmatrix} 0 & 1 & 0 & 0 \\
0 & 0 & 0 & 1 \end{bmatrix}$: matrix specifying the hypotheses on all responses.
 
\item $\boldsymbol{c} = \begin{bmatrix} 0 \\ 0 \end{bmatrix}$: matrix with the values under the null hypothesis. 

\end{itemize}

Thus, the product $\boldsymbol{L}\boldsymbol{\beta}$ represents the initially postulated hypothesis of interest. In this case, the asymptotic distribution of the test is $\chi^2_2$. This specification is very convenient for generating analysis of variance tables and all procedures are easily generalized when there is interest in evaluating hypotheses about the dispersion parameters.


\subsection{ANOVA and MANOVA via Wald test}

Based on the Wald statistics adapted for McGLMs, \citet{htmcglm_proposta} proposed three different procedures for generating ANOVA and MANOVA tables for regression parameters, and a procedure similar to ANOVA and MANOVA to evaluate the dispersion parameters of a model. In the case of ANOVAs, a table is generated for each response variable. For MANOVAs only one table is generated, therefore, in order to be able to perform MANOVAs, the responses must be subject to the same linear predictor.

For purposes of illustration, consider the situation where the goal is to investigate whether two numeric variables denoted by $X_1$ and $X_2$ have an effect on two response variables denoted by $Y_1$ and $Y_2$. For this case, consider the following linear predictor:

$$
g_r(\mu_r) = \beta_{r0} + \beta_{r1} X_1 + \beta_{r2} x_2 + \beta_{r3} X_1X_2.
$$

\noindent where the index $r$ denotes the response variable, $r = 1,2$; $\beta_{r0}$ represents the intercept; $\beta_{r1}$ a regression coefficient associated with the variable $X_1$, $\beta_{r2}$ a regression coefficient associated with the variable $X_2$ and $\beta_{r3}$ a regression coefficient associated with the interaction between $X_1$ and $X_2$. We assume that the units under study are independent, so each response has only one dispersion parameter $\tau_{r0}$ associated with a matrix $Z_0 = I$. Furthermore, we consider that the power parameters have been fixed.

The type II analysis of variance described in \citet{htmcglm_proposta} tests, on each line, whether the complete model differs from the model without a variable. If there are interactions in the model, the complete model is tested against the model without the main effect and any interaction effect involving the variable. In this way, the effect of that variable on the complete model becomes better interpretable, that is, the impact on the quality of the model if we removed a certain variable. Considering the following linear predictor, the type II analysis of variance would do the following tests:

\begin{enumerate}
  \item Tests if the intercept is equal to $0$.
  
  \item Tests if the parameters referring to $X_1$ are equal to $0$. That is, the impact of removing $X_1$ from the model is evaluated. In this case, the interaction is removed because it contains $X_1$.
  
  \item Tests if the parameters referring to $X_2$ are equal to $0$. That is, the impact of removing $X_2$ from the model is evaluated. In this case, the interaction is removed because it contains $X_2$.
  
  \item Tests if the interaction effect is $0$.

\end{enumerate}

\subsection{Multiple comparisons test via Wald test}

When ANOVA shows a significant effect of a categorical variable, it is usually of interest to assess which of the levels differ from each other. In this case, we use multiple comparison tests. In the literature there are several procedures to perform such tests, many of them described in \citet{hsu1996multiple}.

Such a situation can be evaluated using the Wald test based on the correct specification of the $\boldsymbol{L}$ matrix. Thus, it is possible to evaluate hypotheses about any possible contrast between the levels of a given categorical variable. Therefore, it is possible to use Wald's statistics to perform multiple comparison tests as well.

The procedure is basically based on three steps: (i) obtain the matrix of linear combinations of the model parameters that result in the adjusted means; (ii) generate the matrix of contrasts, given by subtracting each pair of lines from the matrix of linear combinations; and (iii) select the lines of interest from this matrix and use them as the Wald test hypothesis specification matrix, instead of the $\boldsymbol{L}$ matrix.
	
For example, suppose there is a response variable $Y$ subject to an explanatory variable $X$ of 4 levels: A, B, C and D. To evaluate the effect of the variable $X$, we fit model with the following linear predictor:

$$g(\mu) = \beta_0 + \beta_1[X=B] + \beta_2[X=C] + \beta_3[X=D].$$

\noindent In this parameterization, the first level of the categorical variable is the reference category and, for the other levels, the change to the reference category is measured; this is called the treatment contrast. In this context $\beta_0$ represents the adjusted mean of level A, while $\beta_1$ represents the difference from A to B, $\beta_2$ represents the difference from A to C and $\beta_3$ represents the difference from A to D. With this parameterization it is possible to obtain the predicted value for any of the categories in such a way that if the individual belongs to category A, $\beta_0$ represents the predicted value; if the individual belongs to category B, $\beta_0 + \beta_1$ represents the predicted value; for category C, $\beta_0 + \beta_2$ represents the predicted value, and finally, for category D, $\beta_0 + \beta_3$ represents the predicted value.

In the matrix, these results can be described as:

$$
    \boldsymbol{K_0} = 
      \begin{matrix}
        A\\ 
        B\\ 
        C\\ 
        D 
      \end{matrix} 
    \begin{bmatrix}
      1 & 0 & 0 & 0\\ 
      1 & 1 & 0 & 0\\ 
      1 & 0 & 1 & 0\\ 
      1 & 0 & 0 & 1 
    \end{bmatrix}
$$

Note that the product $\boldsymbol{K_0} \boldsymbol{\beta}$ generates the vector of predictions for each level of $X$. Thus, we subtract the rows from the matrix of linear combinations $\boldsymbol{K_0}$ in order to generate a matrix of contrasts $\boldsymbol{K_1}$ as in the following:

$$
    \boldsymbol{K_1} = 
      \begin{matrix}
        A-B\\ 
        A-C\\ 
        A-D\\ 
        B-C\\
        B-D\\
        C-D\\ 
      \end{matrix} 
    \begin{bmatrix}
      0 & -1 &  0 &  0\\ 
      0 &  0 & -1 &  0\\ 
      0 &  0 &  0 & -1\\ 
      0 &  1 & -1 &  0\\ 
      0 &  1 &  0 & -1\\ 
      0 &  0 &  1 & -1 
    \end{bmatrix}
$$

To carry out a test of multiple comparisons, we just select the desired contrasts in the lines of the matrix $\boldsymbol{K_1}$ and use these lines as a matrix for specifying the hypotheses of the Wald test. Finally, as usual in tests of multiple comparisons, correction of p-values by means of Bonferroni correction is recommended.

It is important to emphasize that to carry out this procedure for McGLMs, we have class of multivariate models. Thus, as in the case of analysis of variance, for tests of multiple comparisons there are two possibilities: tests for a single response and tests for multiple responses.

In practice, if the interest is a multivariate multiple comparison test, there is a need for all responses to be subject to the same linear predictor and it is enough to expand the contrast matrix using the Kronecker product. In the case of a multiple comparison test for each response, simply select the vector of estimates and the partition corresponding to the matrix $J_{\boldsymbol{\theta}}^{-1}$ for the specific response and proceed with the test as usual.


\section{Implementation}\label{sec:implementacao}

The functions implemented in the package \pkg{htmcglm} generate results showing degrees of freedom and p-values based on the Wald test applied to an object of the \code{mcglm} class. Table \autoref{tab:funcoes} shows the names and a brief description of the implemented functions.

\begin{table}[h]
\centering
\begin{tabular}{ll}
\hline
function                   & Description \\ 
\hline

\code{mc_anova_I()}           & ANOVA type I \\
\code{mc_anova_II()}          & ANOVA type II \\
\code{mc_anova_III()}         & ANOVA type III \\

\code{mc_manova_I()}          & MANOVA type I \\
\code{mc_manova_II()}         & MANOVA type II \\
\code{mc_manova_III()}        & MANOVA type III \\

\code{mc_anova_dispersion()}  & ANOVA type III for dispersion \\
\code{mc_manova_dispersion()} & MANOVA type III for dispersion \\

\code{mc_multcomp()}          & Multiple comparison tests per response \\

\code{mc_mult_multcomp()}     & Multivariate multiple comparison tests \\

\code{mc_linear_hypothesis()} & User-specified general linear hypothesis \\

\hline
\end{tabular}
\caption{Functions implemented in the htmcglm package.}
\label{tab:funcoes}
\end{table}

The functions \code{mc_anova_I()}, \code{mc_anova_II()} and \code{mc_anova_III()} are functions designed to evaluate regression parameters; they generate analysis of variance tables per response for an object of the \code{mcglm} class. The functions \code{mc_manova_I()}, \code{mc_manova_II()} and \code{mc_manova_III()} are also functions designed to evaluate the regression parameters of the model; they generate multivariate analysis of variance tables for a McGLM where the responses are subject to the same predictor. While univariate analysis of variance functions aim to assess the effect of variables for each response, multivariate ones aim to assess the effect of explanatory variables on all response variables simultaneously. The nomenclatures follow what is presented in \citet{htmcglm_proposta} and the functions receive as an argument only the object that stores the fitted model.

As described in \autoref{sec:inf}, the $\boldsymbol{\Omega({\tau})}$ matrix aims to model the correlation between rows of the data set through the so-called matrix linear predictor. In practice, we have for each matrix of the matrix linear predictor an associated dispersion parameter $\tau_d$. Similar to what is done for the mean linear predictor, we can use these estimates to assess the effect of different correlation structures. For this, we implement the functions \code{mc_anova_dispersion()} and \code{mc_manova_dispersion()}.

The \code{mc_anova_dispersion()} function performs an analysis of variance for the model's dispersion parameters. Similar to the other functions with the prefix \code{mc_anova}, a table is generated for each response variable, that is, in the most general cases, we evaluate whether there is evidence that allows us to say that a given dispersion parameter is equal to 0, that is, whether there is an effect of the correlated measures as specified in the matrix linear predictor for that response. The function receives as argument the object in which the model is stored, a list of indices indicating how the dispersion parameters must be tested for each response, in such a way that the dispersion parameters that must be tested together share the same index; the last argument is the set of names to be shown in the final table.

The \code{mc_manova_dispersion()} function can be used in a multivariate model in which the matrix linear predictors are the same for all responses and there is an interest in evaluating whether the effect of correlated measures is the same for all responses. This function receives as argument the object in which the model is stored, a vector of indices indicating how the dispersion parameters must be tested, in such a way that dispersion parameters that must be tested together share the same index; the last argument is the set of names to be shown in the final table.

For  multiple comparisons tests, the functions \code{mc_multcomp()} and \code{mc_mult_multcomp()} were implemented. These functions should be used as a complement to the analysis of variance and multivariate analysis of variance functions when they show a significant effect of categorical explanatory variables. The functions for multiple comparisons are used to perform two-by-two comparisons and identify which levels differ from each other. These functions receive as an argument the model, the variable or variables in which there is interest in evaluating comparisons between levels and also the data used to fit the model.

Finally, the \code{mc_linear_hypothesis()} function is the most flexible one. The \code{mc_linear_hypothesis()} specifies any type of hypothesis about regression, dispersion or power parameters of a McGLM. It is also possible to specify hypotheses on multiple parameters and the vector of null hypothesis values is user defined. This function receives as arguments the model, a vector containing the parameters to be tested and the values under the null hypothesis. With some work, using the general linear hypotheses function, it is possible to replicate the results obtained by the analysis of variance functions.


\section{Examples}\label{sec:exemplos}

In this section we shall provide practical examples of using the functions implemented in the \pkg{htmcglm} package based on multivariate models fitted with the \pkg{mcglm} package.

\subsection{Example 1: soya}

The data are from an experiment carried out in a greenhouse with soybeans. The experimental design has two plants per plot in which each unit was subjected to different combinations of water and fertilizer. There are three levels of a factor corresponding to the amount of water in the soil (\code{water}) and five levels of potassium fertilization (\code{pot}). In addition, the plots were arranged in five blocks (\code{block}). Three response variables were evaluated: grain yield (\code{grain}), number of seeds (\code{seeds}) and number of viable peas per plant (\code{viablepeas}).

This is an interesting dataset to exemplify the use of the implemented functions because there are three response variables of different types: grain yield is a continuous variable, the number of seeds is a count, and the number of viable peas per plant is an example of a binomial variable. The dataset is available in the \code{mcglm} package.

\begin{knitrout}
\definecolor{shadecolor}{rgb}{0.969, 0.969, 0.969}\color{fgcolor}\begin{kframe}
\begin{alltt}
\hlkwd{data}\hlstd{(}\hlstr{"soya"}\hlstd{,} \hlkwc{package} \hlstd{=} \hlstr{"mcglm"}\hlstd{)}
\end{alltt}
\end{kframe}
\end{knitrout}

The objective of the analysis is to evaluate the effect of fertilization and water on the three response variables of interest. For the purposes of analysis, we considered as explanatory variables the levels of water, fertilization and also the interactions between these two factors. Additionally, the block effect was added to the predictors. To fit the model, the first step is to specify the linear predictors.

\begin{knitrout}
\definecolor{shadecolor}{rgb}{0.969, 0.969, 0.969}\color{fgcolor}\begin{kframe}
\begin{alltt}
\hlstd{form.grain} \hlkwb{<-} \hlstd{grain} \hlopt{~} \hlstd{block} \hlopt{+} \hlstd{water} \hlopt{*} \hlstd{pot}
\hlstd{form.seed} \hlkwb{<-} \hlstd{seeds} \hlopt{~} \hlstd{block} \hlopt{+} \hlstd{water} \hlopt{*} \hlstd{pot}

\hlstd{soya}\hlopt{$}\hlstd{viablepeasP} \hlkwb{<-} \hlstd{soya}\hlopt{$}\hlstd{viablepeas} \hlopt{/} \hlstd{soya}\hlopt{$}\hlstd{totalpeas}
\hlstd{form.peas} \hlkwb{<-} \hlstd{viablepeasP} \hlopt{~} \hlstd{block} \hlopt{+} \hlstd{water} \hlopt{*} \hlstd{pot}
\end{alltt}
\end{kframe}
\end{knitrout}

The second step is to specify the matrix linear predictor. We consider in this case that the observations are independent, so we include only one identity matrix.

\begin{knitrout}
\definecolor{shadecolor}{rgb}{0.969, 0.969, 0.969}\color{fgcolor}\begin{kframe}
\begin{alltt}
\hlstd{Z0} \hlkwb{<-} \hlkwd{mc_id}\hlstd{(soya)}
\end{alltt}
\end{kframe}
\end{knitrout}

With the elements defined, we can fit the model. Through the function \code{mcglm()} we specify the linear predictors for the mean, the matrices of the matrix linear predictors, the link and variance functions, the number of trials for the binomial variable and whether or not we are interested in estimating the power parameters. For more details on specifying predictors and fit McGLMs, see \citet{Bonat16} and \citet{mcglm}.

\begin{knitrout}
\definecolor{shadecolor}{rgb}{0.969, 0.969, 0.969}\color{fgcolor}\begin{kframe}
\begin{alltt}
\hlstd{fit_joint} \hlkwb{<-} \hlkwd{mcglm}\hlstd{(}\hlkwc{linear_pred} \hlstd{=} \hlkwd{c}\hlstd{(form.grain,}
                                   \hlstd{form.seed,}
                                   \hlstd{form.peas),}
                   \hlkwc{matrix_pred} \hlstd{=} \hlkwd{list}\hlstd{(}\hlkwd{c}\hlstd{(Z0),}
                                      \hlkwd{c}\hlstd{(Z0),}
                                      \hlkwd{c}\hlstd{(Z0)),}
                   \hlkwc{link} \hlstd{=} \hlkwd{c}\hlstd{(}\hlstr{"identity"}\hlstd{,}
                            \hlstr{"log"}\hlstd{,}
                            \hlstr{"logit"}\hlstd{),}
                   \hlkwc{variance} \hlstd{=} \hlkwd{c}\hlstd{(}\hlstr{"constant"}\hlstd{,}
                                \hlstr{"tweedie"}\hlstd{,}
                                \hlstr{"binomialP"}\hlstd{),}
                   \hlkwc{Ntrial} \hlstd{=} \hlkwd{list}\hlstd{(}\hlkwa{NULL}\hlstd{,}
                                 \hlkwa{NULL}\hlstd{,}
                                 \hlstd{soya}\hlopt{$}\hlstd{totalpeas),}
                   \hlkwc{power_fixed} \hlstd{=} \hlkwd{c}\hlstd{(T,T,T),}
                   \hlkwc{data} \hlstd{= soya)}
\end{alltt}
\end{kframe}
\end{knitrout}

To evaluate some results of the model it is possible to use the function \code{summary()} that returns the formula of the linear predictors, the link, variance and covariance functions specified to fit the model, the estimates of the regression and dispersion parameters as well as standard errors.

With the fitted model, we can apply the implemented functions to evaluate the regression and dispersion parameters of the model. The analysis of variance functions depend only on the object that contains the fitted model and return a table for each response.

\subsubsection{ANOVA type I}
 
\begin{knitrout}
\definecolor{shadecolor}{rgb}{0.969, 0.969, 0.969}\color{fgcolor}\begin{kframe}
\begin{alltt}
\hlkwd{mc_anova_I}\hlstd{(fit_joint)}
\end{alltt}
\begin{verbatim}
## ANOVA type I using Wald statistic for fixed effects
## 
## Call: grain ~ block + water * pot
## 
##   Covariate Df       Chi Pr(>Chi)
## 1 Intercept 19 6283.6472    0e+00
## 2     block 18  419.6702    0e+00
## 3     water 14  405.1498    0e+00
## 4       pot 12  350.9316    0e+00
## 5 water:pot  8   30.4494    2e-04
## 
## Call: seeds ~ block + water * pot
## 
##   Covariate Df         Chi Pr(>Chi)
## 1 Intercept 19 127429.2620   0.0000
## 2     block 18    205.8174   0.0000
## 3     water 14    194.0161   0.0000
## 4       pot 12    130.2022   0.0000
## 5 water:pot  8     12.7366   0.1212
## 
## Call: viablepeasP ~ block + water * pot
## 
##   Covariate Df      Chi Pr(>Chi)
## 1 Intercept 19 971.1096   0.0000
## 2     block 18 300.2990   0.0000
## 3     water 14 297.4306   0.0000
## 4       pot 12 295.2420   0.0000
## 5 water:pot  8  20.0549   0.0101
\end{verbatim}
\end{kframe}
\end{knitrout}

\subsubsection{ANOVA type II}

\begin{knitrout}
\definecolor{shadecolor}{rgb}{0.969, 0.969, 0.969}\color{fgcolor}\begin{kframe}
\begin{alltt}
\hlkwd{mc_anova_II}\hlstd{(fit_joint)}
\end{alltt}
\begin{verbatim}
## ANOVA type II using Wald statistic for fixed effects
## 
## Call: grain ~ block + water * pot
## 
##   Covariate Df      Chi Pr(>Chi)
## 1 Intercept  1 102.2961   0.0000
## 2     block  4  14.3051   0.0064
## 3     water 10  84.6677   0.0000
## 4       pot 12 350.9316   0.0000
## 5 water:pot  8  30.4494   0.0002
## 
## Call: seeds ~ block + water * pot
## 
##   Covariate Df       Chi Pr(>Chi)
## 1 Intercept  1 3993.9442   0.0000
## 2     block  4   11.6363   0.0203
## 3     water 10   70.8041   0.0000
## 4       pot 12  130.2022   0.0000
## 5 water:pot  8   12.7366   0.1212
## 
## Call: viablepeasP ~ block + water * pot
## 
##   Covariate Df      Chi Pr(>Chi)
## 1 Intercept  1  13.4353   0.0002
## 2     block  4   4.4305   0.3509
## 3     water 10  33.9928   0.0002
## 4       pot 12 295.2420   0.0000
## 5 water:pot  8  20.0549   0.0101
\end{verbatim}
\end{kframe}
\end{knitrout}

\subsubsection{ANOVA type III}

\begin{knitrout}
\definecolor{shadecolor}{rgb}{0.969, 0.969, 0.969}\color{fgcolor}\begin{kframe}
\begin{alltt}
\hlkwd{mc_anova_III}\hlstd{(fit_joint)}
\end{alltt}
\begin{verbatim}
## ANOVA type III using Wald statistic for fixed effects
## 
## Call: grain ~ block + water * pot
## 
##   Covariate Df      Chi Pr(>Chi)
## 1 Intercept  1 102.2961   0.0000
## 2     block  4  14.3051   0.0064
## 3     water  2   2.3991   0.3013
## 4       pot  4  64.0038   0.0000
## 5 water:pot  8  30.4494   0.0002
## 
## Call: seeds ~ block + water * pot
## 
##   Covariate Df       Chi Pr(>Chi)
## 1 Intercept  1 3993.9442   0.0000
## 2     block  4   11.6363   0.0203
## 3     water  2    3.9399   0.1395
## 4       pot  4   19.1997   0.0007
## 5 water:pot  8   12.7366   0.1212
## 
## Call: viablepeasP ~ block + water * pot
## 
##   Covariate Df     Chi Pr(>Chi)
## 1 Intercept  1 13.4353   0.0002
## 2     block  4  4.4305   0.3509
## 3     water  2  5.2513   0.0724
## 4       pot  4 71.1026   0.0000
## 5 water:pot  8 20.0549   0.0101
\end{verbatim}
\end{kframe}
\end{knitrout}

Similarly, multivariate analysis of variance functions also depend only on the fitted model. It is important to note that for practical purposes the multivariate analysis of variance functions require the predictors for all responses to be the same.

\subsubsection{MANOVA type I}

\begin{knitrout}
\definecolor{shadecolor}{rgb}{0.969, 0.969, 0.969}\color{fgcolor}\begin{kframe}
\begin{alltt}
\hlkwd{mc_manova_I}\hlstd{(fit_joint)}
\end{alltt}
\begin{verbatim}
## MANOVA type I using Wald statistic for fixed effects
## 
## Call: ~ block+water*pot
##   Covariate Df         Chi Pr(>Chi)
## 1 Intercept 57 168255.3139        0
## 2     block 54    816.7633        0
## 3     water 42    794.0601        0
## 4       pot 36    708.8164        0
## 5 water:pot 24     68.7879        0
\end{verbatim}
\end{kframe}
\end{knitrout}

\subsubsection{MANOVA type II}

\begin{knitrout}
\definecolor{shadecolor}{rgb}{0.969, 0.969, 0.969}\color{fgcolor}\begin{kframe}
\begin{alltt}
\hlkwd{mc_manova_II}\hlstd{(fit_joint)}
\end{alltt}
\begin{verbatim}
## MANOVA type II using Wald statistic for fixed effects
## 
## Call: ~ block+water*pot
##   Covariate Df       Chi Pr(>Chi)
## 1 Intercept  3 5553.7954    0.000
## 2     block 12   23.7478    0.022
## 3     water 30  160.9564    0.000
## 4       pot 36  708.8164    0.000
## 5 water:pot 24   68.7879    0.000
\end{verbatim}
\end{kframe}
\end{knitrout}

\subsubsection{MANOVA type III}

\begin{knitrout}
\definecolor{shadecolor}{rgb}{0.969, 0.969, 0.969}\color{fgcolor}\begin{kframe}
\begin{alltt}
\hlkwd{mc_manova_III}\hlstd{(fit_joint)}
\end{alltt}
\begin{verbatim}
## MANOVA type III using Wald statistic for fixed effects
## 
## Call: ~ block+water*pot
##   Covariate Df       Chi Pr(>Chi)
## 1 Intercept  3 5553.7954   0.0000
## 2     block 12   23.7478   0.0220
## 3     water  6    9.0173   0.1726
## 4       pot 12  149.0321   0.0000
## 5 water:pot 24   68.7879   0.0000
\end{verbatim}
\end{kframe}
\end{knitrout}

For general linear hypotheses about regression parameters, it is sufficient to specify the model and the hypothesis to be tested. To identify the parameters of interest, use the \code{coef()} function.

\subsubsection{Test on a single regression parameter}

\begin{knitrout}
\definecolor{shadecolor}{rgb}{0.969, 0.969, 0.969}\color{fgcolor}\begin{kframe}
\begin{alltt}
\hlkwd{mc_linear_hypothesis}\hlstd{(}\hlkwc{object} \hlstd{=  fit_joint,}
                     \hlkwc{hypothesis} \hlstd{=} \hlkwd{c}\hlstd{(}\hlstr{'beta11 = 0'}\hlstd{))}
\end{alltt}
\begin{verbatim}
## Linear hypothesis test
## 
## Hypothesis:            
## 1 beta11 = 0
## 
## Results:
##   Df    Chi Pr(>Chi)
## 1  1 1.2362   0.2662
\end{verbatim}
\end{kframe}
\end{knitrout}

\subsubsection{Test on more than one regression parameter}

\begin{knitrout}
\definecolor{shadecolor}{rgb}{0.969, 0.969, 0.969}\color{fgcolor}\begin{kframe}
\begin{alltt}
\hlkwd{mc_linear_hypothesis}\hlstd{(}\hlkwc{object} \hlstd{=  fit_joint,}
                     \hlkwc{hypothesis} \hlstd{=} \hlkwd{c}\hlstd{(}\hlstr{'beta11 = 0'}\hlstd{,}
                                    \hlstr{'beta12 = 0'}\hlstd{))}
\end{alltt}
\begin{verbatim}
## Linear hypothesis test
## 
## Hypothesis:            
## 1 beta11 = 0
## 2 beta12 = 0
## 
## Results:
##   Df    Chi Pr(>Chi)
## 1  2 3.5639   0.1683
\end{verbatim}
\end{kframe}
\end{knitrout}

\subsubsection{Test of equality of effects between regression parameters}

\begin{knitrout}
\definecolor{shadecolor}{rgb}{0.969, 0.969, 0.969}\color{fgcolor}\begin{kframe}
\begin{alltt}
\hlkwd{mc_linear_hypothesis}\hlstd{(}\hlkwc{object} \hlstd{=  fit_joint,}
                     \hlkwc{hypothesis} \hlstd{=} \hlkwd{c}\hlstd{(}\hlstr{'beta11 = beta21'}\hlstd{))}
\end{alltt}
\begin{verbatim}
## Linear hypothesis test
## 
## Hypothesis:                 
## 1 beta11 = beta21
## 
## Results:
##   Df    Chi Pr(>Chi)
## 1  1 1.3491   0.2454
\end{verbatim}
\end{kframe}
\end{knitrout}

\subsection{Example 2: Hunting}

The Hunting dataset, presented in \citet{hunting}, this dataset is also available in the package \pkg{mcglm}. The data addresses a problem where responses are longitudinal bivariate counts on animals hunted in Basile Fang village, Bioko North Province, Bioko Island, Equatorial Guinea. The response variables are: monthly numbers of blue duikers (\code{BD}) and other small animals (\code{OT}) shot or captured in a random sample of 52 commercial hunters from August 2010 to September 2013. Assume that the interest is to evaluate the effect of a factor with 2 levels that indicates if the animal was hunted by means of a firearm or trap (\code{METHOD}) and a factor with 2 levels that indicates the sex of the animal (\code{ SEX}).

\begin{knitrout}
\definecolor{shadecolor}{rgb}{0.969, 0.969, 0.969}\color{fgcolor}\begin{kframe}
\begin{alltt}
\hlkwd{data}\hlstd{(}\hlstr{"Hunting"}\hlstd{,} \hlkwc{package} \hlstd{=} \hlstr{"mcglm"}\hlstd{)}
\end{alltt}
\end{kframe}
\end{knitrout}

As in the first example, to fit the model it is necessary to define the linear predictors for the mean, the matrices of the linear matrix predictors, the link and variance functions, whether or not we are interested in estimating the power parameters. For this analysis, we considered in the matrix predictor the structure of repeated measures introduced by the observations taken for the same hunter and month (\code{HUNTER.MONTH}) and the number of hunting days per month was used as an offset term.

\begin{knitrout}
\definecolor{shadecolor}{rgb}{0.969, 0.969, 0.969}\color{fgcolor}\begin{kframe}
\begin{alltt}
\hlstd{form.OT} \hlkwb{<-} \hlstd{OT} \hlopt{~} \hlstd{METHOD} \hlopt{*} \hlstd{SEX}
\hlstd{form.BD} \hlkwb{<-} \hlstd{BD} \hlopt{~} \hlstd{METHOD} \hlopt{*} \hlstd{SEX}

\hlstd{Z0} \hlkwb{<-} \hlkwd{mc_id}\hlstd{(Hunting)}
\hlstd{Z1} \hlkwb{<-} \hlkwd{mc_mixed}\hlstd{(}\hlopt{~} \hlnum{0} \hlopt{+} \hlstd{HUNTER.MONTH,} \hlkwc{data} \hlstd{= Hunting)}

\hlstd{fit} \hlkwb{<-} \hlkwd{mcglm}\hlstd{(}\hlkwc{linear_pred} \hlstd{=} \hlkwd{c}\hlstd{(form.BD, form.OT),}
             \hlkwc{matrix_pred} \hlstd{=} \hlkwd{list}\hlstd{(}\hlkwd{c}\hlstd{(Z0, Z1),}
                                \hlkwd{c}\hlstd{(Z0, Z1)),}
             \hlkwc{link} \hlstd{=} \hlkwd{c}\hlstd{(}\hlstr{"log"}\hlstd{,} \hlstr{"log"}\hlstd{),}
             \hlkwc{variance} \hlstd{=} \hlkwd{c}\hlstd{(}\hlstr{"poisson_tweedie"}\hlstd{,}
                          \hlstr{"poisson_tweedie"}\hlstd{),}
             \hlkwc{offset} \hlstd{=} \hlkwd{list}\hlstd{(}\hlkwd{log}\hlstd{(Hunting}\hlopt{$}\hlstd{OFFSET),}
                           \hlkwd{log}\hlstd{(Hunting}\hlopt{$}\hlstd{OFFSET)),}
             \hlkwc{data} \hlstd{= Hunting)}
\end{alltt}
\end{kframe}
\end{knitrout}

Again, to evaluate some model results it is possible to use the \code{summary()} function. We can also apply the already presented functions implemented for ANOVAs, MANOVAs and tests of general linear hypotheses on the regression parameters of the model.

In this case, as there is a specified matrix linear predictor, an in-depth study of the dispersion parameters may be of interest. This analysis can be done with the already used function \code{mc_linear_hypothesis()}.

\subsubsection{Test on a single dispersion parameter}

\begin{knitrout}
\definecolor{shadecolor}{rgb}{0.969, 0.969, 0.969}\color{fgcolor}\begin{kframe}
\begin{alltt}
\hlkwd{mc_linear_hypothesis}\hlstd{(}\hlkwc{object} \hlstd{=  fit,}
                     \hlkwc{hypothesis} \hlstd{=} \hlkwd{c}\hlstd{(}\hlstr{'tau11 = 0'}\hlstd{))}
\end{alltt}
\begin{verbatim}
## Linear hypothesis test
## 
## Hypothesis:           
## 1 tau11 = 0
## 
## Results:
##   Df     Chi Pr(>Chi)
## 1  1 22.5613        0
\end{verbatim}
\end{kframe}
\end{knitrout}

\subsubsection{Test on more than one dispersion parameter}

\begin{knitrout}
\definecolor{shadecolor}{rgb}{0.969, 0.969, 0.969}\color{fgcolor}\begin{kframe}
\begin{alltt}
\hlkwd{mc_linear_hypothesis}\hlstd{(}\hlkwc{object} \hlstd{=  fit,}
                     \hlkwc{hypothesis} \hlstd{=} \hlkwd{c}\hlstd{(}\hlstr{'tau11 = 0'}\hlstd{,}
                                    \hlstr{'tau21 = 0'}\hlstd{))}
\end{alltt}
\begin{verbatim}
## Linear hypothesis test
## 
## Hypothesis:           
## 1 tau11 = 0
## 2 tau21 = 0
## 
## Results:
##   Df    Chi Pr(>Chi)
## 1  2 29.098        0
\end{verbatim}
\end{kframe}
\end{knitrout}

\subsubsection{Test of equality of effects between dispersion parameters}

\begin{knitrout}
\definecolor{shadecolor}{rgb}{0.969, 0.969, 0.969}\color{fgcolor}\begin{kframe}
\begin{alltt}
\hlkwd{mc_linear_hypothesis}\hlstd{(}\hlkwc{object} \hlstd{=  fit,}
                     \hlkwc{hypothesis} \hlstd{=} \hlkwd{c}\hlstd{(}\hlstr{'tau12 = tau22'}\hlstd{))}
\end{alltt}
\begin{verbatim}
## Linear hypothesis test
## 
## Hypothesis:               
## 1 tau12 = tau22
## 
## Results:
##   Df    Chi Pr(>Chi)
## 1  1 5.8183   0.0159
\end{verbatim}
\end{kframe}
\end{knitrout}

To evaluate dispersion parameters, we have the procedure analogous to the analysis of variance for regression parameters. These functions require specifying more arguments: one that determines the relationship between dispersion parameters and the other that specifies the names that will appear in the final output.

\subsubsection{ANOVA type III for dispersion}

\begin{knitrout}
\definecolor{shadecolor}{rgb}{0.969, 0.969, 0.969}\color{fgcolor}\begin{kframe}
\begin{alltt}
\hlkwd{mc_anova_dispersion}\hlstd{(fit,}
                    \hlkwc{p_var} \hlstd{=} \hlkwd{list}\hlstd{(}\hlkwd{c}\hlstd{(}\hlnum{0}\hlstd{,}\hlnum{1}\hlstd{),} \hlkwd{c}\hlstd{(}\hlnum{0}\hlstd{,}\hlnum{1}\hlstd{)),}
                    \hlkwc{names} \hlstd{=} \hlkwd{list}\hlstd{(}\hlkwd{c}\hlstd{(}\hlstr{'tau10'}\hlstd{,} \hlstr{'tau11'}\hlstd{),}
                                 \hlkwd{c}\hlstd{(}\hlstr{'tau20'}\hlstd{,} \hlstr{'tau21'}\hlstd{)))}
\end{alltt}
\begin{verbatim}
## ANOVA type III using Wald statistic for dispersion parameters
## 
## Call: BD ~ METHOD * SEX
## 
##   Dispersion Df     Chi Pr(>Chi)
## 1      tau10  1 22.5613        0
## 2      tau11  1 97.0998        0
## 
## Call: OT ~ METHOD * SEX
## 
##   Dispersion Df     Chi Pr(>Chi)
## 1      tau20  1  7.2008   0.0073
## 2      tau21  1 29.0133   0.0000
\end{verbatim}
\end{kframe}
\end{knitrout}

\subsubsection{MANOVA type III for dispersion}

\begin{knitrout}
\definecolor{shadecolor}{rgb}{0.969, 0.969, 0.969}\color{fgcolor}\begin{kframe}
\begin{alltt}
\hlkwd{mc_manova_dispersion}\hlstd{(fit,}
                     \hlkwc{p_var} \hlstd{=} \hlkwd{c}\hlstd{(}\hlnum{0}\hlstd{,}\hlnum{1}\hlstd{),}
                     \hlkwc{names} \hlstd{=} \hlkwd{c}\hlstd{(}\hlstr{'tau0'}\hlstd{,} \hlstr{'tau1'}\hlstd{))}
\end{alltt}
\begin{verbatim}
## MANOVA type III using Wald statistic for dispersion parameters
## 
## Call: ~ METHOD*SEX
##   Covariate Df      Chi Pr(>Chi)
## 1      tau0  2  29.0980        0
## 2      tau1  2 124.2049        0
\end{verbatim}
\end{kframe}
\end{knitrout}

Finally, we can use the functions for testing multiple comparisons to assess differences between levels of categorical explanatory variables included in the model. 

\subsubsection{Univariate multiple comparisons test}

\begin{knitrout}
\definecolor{shadecolor}{rgb}{0.969, 0.969, 0.969}\color{fgcolor}\begin{kframe}
\begin{alltt}
\hlkwd{mc_multcomp}\hlstd{(}\hlkwc{object} \hlstd{= fit,}
            \hlkwc{effect} \hlstd{=} \hlkwd{list}\hlstd{(}\hlkwd{c}\hlstd{(}\hlstr{'METHOD'}\hlstd{,} \hlstr{'SEX'}\hlstd{),}
                          \hlkwd{c}\hlstd{(}\hlstr{'METHOD'}\hlstd{,} \hlstr{'SEX'}\hlstd{)),}
            \hlkwc{data} \hlstd{= Hunting)}
\end{alltt}
\begin{verbatim}
## Multiple comparisons test for each outcome using Wald statistic
## 
## Call: BD ~ METHOD * SEX
## 
##                        Contrast Df      Chi Pr(>Chi)
## 1 Escopeta:Female-Escopeta:Male  1 175.7657        0
## 2 Escopeta:Female-Trampa:Female  1  20.1379        0
## 3   Escopeta:Female-Trampa:Male  1  35.6372        0
## 4     Escopeta:Male-Trampa:Male  1  24.3946        0
## 5   Trampa:Female-Escopeta:Male  1 217.7398        0
## 6     Trampa:Female-Trampa:Male  1 132.6125        0
## 
## Call: OT ~ METHOD * SEX
## 
##                        Contrast Df     Chi Pr(>Chi)
## 1 Escopeta:Female-Escopeta:Male  1 14.3969   0.0009
## 2 Escopeta:Female-Trampa:Female  1  6.5843   0.0617
## 3   Escopeta:Female-Trampa:Male  1  5.6455   0.1050
## 4     Escopeta:Male-Trampa:Male  1  0.7480   1.0000
## 5   Trampa:Female-Escopeta:Male  1 31.3069   0.0000
## 6     Trampa:Female-Trampa:Male  1 25.3203   0.0000
\end{verbatim}
\end{kframe}
\end{knitrout}

\subsubsection{Multivariate multiple comparisons test}

\begin{knitrout}
\definecolor{shadecolor}{rgb}{0.969, 0.969, 0.969}\color{fgcolor}\begin{kframe}
\begin{alltt}
\hlkwd{mc_mult_multcomp}\hlstd{(}\hlkwc{object} \hlstd{= fit,}
                 \hlkwc{effect} \hlstd{=} \hlkwd{c}\hlstd{(}\hlstr{'METHOD'}\hlstd{,} \hlstr{'SEX'}\hlstd{),}
                 \hlkwc{data} \hlstd{= Hunting)}
\end{alltt}
\begin{verbatim}
## Multivariate multiple comparisons test using Wald statistic
## 
## Call: ~ METHOD*SEX
##                        Contrast Df      Chi Pr(>Chi)
## 1 Escopeta:Female-Escopeta:Male  2 215.0490        0
## 2 Escopeta:Female-Trampa:Female  2  31.8503        0
## 3   Escopeta:Female-Trampa:Male  2  47.8804        0
## 4     Escopeta:Male-Trampa:Male  2  27.5459        0
## 5   Trampa:Female-Escopeta:Male  2 287.6161        0
## 6     Trampa:Female-Trampa:Male  2 184.8844        0
\end{verbatim}
\end{kframe}
\end{knitrout}


\section{Concluding remarks}\label{sec:conclusao}

This article described the \proglang{R} implementation of procedures to perform hypothesis tests on McGLMs parameters based on Wald statistics. McGLMs have regression, dispersion, power and correlation parameters; each set of parameters has a very relevant practical interpretation in the context of problem analysis with potential multiple responses as a function of a set of explanatory variables.

Based on the proposed use of the Wald test for McGLMs, we developed the \pkg{htmcglm} with procedures for testing general linear hypotheses, generating ANOVA and MANOVA tables for regression and dispersion parameters and also multiple comparisons tests. All these procedures were implemented in the \proglang{R} language and complement the existing functionalities in the \pkg{mcglm} library.

The discussed examples illustrate how to evaluate the most common hypotheses that arise in regression problems: evaluating parameters individually and evaluating sets of parameters. We focused our efforts on tools to evaluate regression and dispersion parameters, because by studying regression parameters it is possible to identify the variables that have a significant effect on the response; on the other hand, the dispersion parameters allow assessing whether there is an effect of correlated observations. In this way, the study of these quantities provides valuable information about the importance of the elements of a multivariate regression model.

Possible extensions of the \pkg{htmcglm} package follow the idea of evaluation of McGLMs parameters for a better understanding of the impact of elements in modeling problems. Some possibilities are: exploring corrections of p-values according to the size of the tested hypotheses, exploring procedures beyond the Wald test (such as the Score test and the pseudo likelihood ratio test), implementing new procedures for multiple comparisons, adapting the proposal to deal with alternative contrasts to the usual ones, explore procedures for automatic selection of covariates (backward elimination, forward selection, stepwise selection) and also covariate selection through the inclusion of penalty in the complexity adjustment (similar to the idea of spline regression).


\section*{Acknowledgments}

\begin{leftbar}
This study was financed in part by the Coordenação de Aperfeiçoamento de Pessoal de Nível Superior – Brasil (CAPES) – Finance Code 001.
\end{leftbar}


\bibliography{refs}


\newpage





\end{document}